\documentclass[letterpaper,superscriptaddress,english,aps,prb,twocolumn,floatfix, showpacs, amsfonts amssymb]{revtex4-1}

\usepackage{epsfig, graphicx,psfrag,amsmath,amssymb,float}
\usepackage[T1]{fontenc}
\usepackage[latin9]{inputenc}
\usepackage{amsmath}
\usepackage{amssymb}
\usepackage{amscd}
\usepackage{bm}
\usepackage{psfrag}
\usepackage{babel}

\newcommand{\be}{\begin{equation}}
\newcommand{\ee}{\end{equation}}
\newcommand{\bea}{\begin{eqnarray}}
\newcommand{\eea}{\end{eqnarray}}

\begin{document}

\title{Majorana Fermions in Strongly Interacting Helical Liquids}

\author{Eran Sela, Alexander Altland, and Achim Rosch}

\affiliation{Institute for Theoretical Physics, University of Cologne, 50937 Cologne, Germany}

\begin{abstract}
Majorana fermions were proposed to occur at edges and interfaces of gapped one-dimensional
systems where phases with different topological character meet due to an
interplay of spin-orbit coupling, proximity-induced superconductivity and external magnetic fields. Here we investigate the effect of strong particle interactions, and show that the helical liquid offers a mechanism that protects the very existence of Majorana edge states: whereas moderate interactions close the proximity gap which supports the edge states, in helical liquids the gap re-opens due to two-particle processes. However, gapless fermionic excitations occur at spatial proximity to the Majorana states at interfaces and may jeopardize their long term Majorana
coherence.
\end{abstract}


\maketitle
\section{Introduction}
Majorana fermions (MFs) were introduced in 1937 as neutral
particles that are their own antiparticles, with the aim of describing neutrinos.~\cite{majorana} More recently an increasing number of candidate MFs were suggested  as
quasiparticles in condensed matter systems, including the Moore-Read quantum Hall state~\cite{moore} at filling fraction $5/2$, which can be described as a $p$ wave superconductor of composite fermions.~\cite{read} In this picture MFs are zero energy bound states localized at vortices~\cite{kopnin} and show non-abelian statistics upon braiding those vortices;~\cite{volovik}  for a recent review see Refs.~\onlinecite{Nayak,wilczec}.

Searching for alternative experimentally feasible candidates, inspired by the one-dimensional (1D) model of Kitaev,~\cite{kitaev01} several groups have proposed
physical realizations of MFs as edge states of 1D systems, including electrostatic defect lines in superconductors,~\cite{wimmer}
semiconductor quantum wires proximity coupled to a superconductor,~\cite{Lutchyng,oreg}
quasi-1D superconductors,~\cite{Potter} and cold atoms trapped in
1D.~\cite{jiang}  In semiconductor wires these edge states can be
controlled by tuning  external gates, and
networks of such wires are envisioned to perform non-abelian quantum
computation.~\cite{alicea,Sau}

\begin{figure}[h] \begin{center} \includegraphics*[width=85mm]{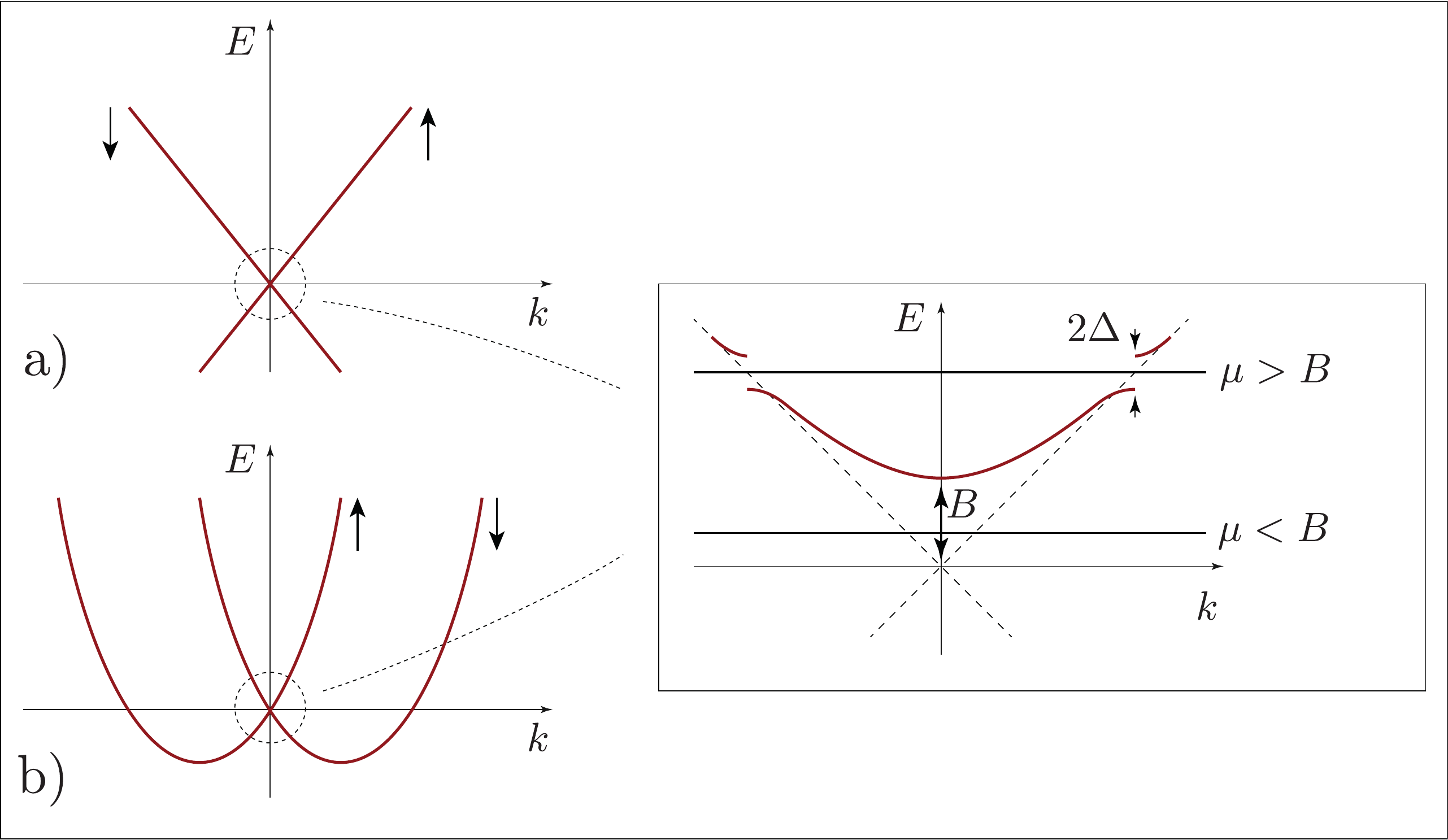}
    \caption{Helical liquid formed by (a) surface bands of a 2D
      topological insulator, or (b) the low-momentum excitations of a
      quantum wire subject Rashba interaction. Inset: Gaps are induced both by magnetic field ($B$) or superconductivity ($\Delta$).
      While for $\mu<\sqrt{ B^2-\Delta^2}, \Delta<B$ and no interactions, the gap is controlled by $B$,
      the superconducting gap dominates for $\mu>\sqrt{B^2-\Delta^2}$ and a different topological sector is obtained.
\label{fg:0}}
\end{center}
\end{figure}

The practicability of such 1D applications critically hinges on the
stability of Majorana fermion states against particle
interactions. Focusing on the most elementary realization of a
quantum wire with a \textit{single} spin polarized fermion band proximity
coupled to a superconductor and with spin-orbit coupling, which was shown~\cite{alicea} to reduce to Kitaev's model~\cite{kitaev01} -- Gangadharaiah et al.~\cite{loss}  indeed have shown that
even moderately strong interaction may compromise the stability of
Majorana fermions: beyond a certain strength, interactions remove the
proximity gap, and along with it any accompanying Majorana bound
states. Here we consider the impact of interactions on the
full `helical liquid', i.e. a system of \textit{two}
counter-propagating fermion bands carrying opposite spin, as realized as the
surface state of two-dimensional topological insulators,~\cite{Konig} or proximity
coupled semiconductor quantum wires subject to spin-orbit interaction,~\cite{Lutchyng,oreg} cf. Fig.~\ref{fg:0}. We show
that Majorana fermion states forming in these systems enjoy a much
higher degree of stability. Conceptually, the resilience of the
helical liquid to interactions is rooted in
momentum conserving two-particle scattering processes between the
constituting fermion bands.~\cite{foonoteloss,Xu,bernevig} Such processes open gaps
that resist interactions and ultimately lead
to the stabilization of Majorana states. At the same time, strong interactions may act as a source of Majorana
fermion \textit{decoherence}.

The outline of the paper is as follows. Having introduced the generic model for an interacting helical liquid in Sec.~\ref{se:model} and mapped it into a spin chain model, we map out
the helical liquid's global phase diagram in Sec.~\ref{se:pd} in a parameter space spanned
by proximity coupling, external magnetic field strength, interaction
strength and chemical potential. At strong interactions the system
turns out to support a phase of gapless fermionic excitations in close \textit{parametric}
neighborhood of the Majorana/Ising quantum critical point.~\cite{zuber} In a
quantum wire subject to slowly changing parameter profiles, considered in Sec.~\ref{se:interfaces}, this means
the existence of a domain of low lying excitations in close
\textit{spatial} proximity to a localized Majorana state. As pointed out in the concluding section \ref{se:discussion}, such
excitations may spoil the long term coherence of the Majorana fermion, however
 fermion parity protected readout schemes may be employed to probe a certain decoherence free subspace.~\cite{akhmerov,Hassler} Details on the derivation of the phase diagram using bosonization methods are included in the appendix.

\section{Model}
\label{se:model}
A helical liquid is defined by the Hamiltonian
$H_0=\int dx\, \mathcal{H}_0(x)$,
\begin{equation}
  \mathcal{H}_0= \psi_{L \downarrow}^\dagger (v_F i \partial_x-\mu) \psi_{L \downarrow} + \psi_{R \uparrow}^\dagger (-v_F i \partial_x-\mu) \psi_{R \uparrow},
\end{equation}
where $\psi_{R \uparrow}$($\psi_{L \downarrow}$) are right (left)
moving spin up (down) fermion fields, $v_F$ is the Fermi velocity, and
$\mu$ the chemical potential.  Gaps of different topological
signature may be realized by coupling the system
to a magnetic field $B$ (Zeeman gap), and to an s-wave superconductor (proximity gap),
\begin{eqnarray}
  \label{BD}
  \delta \mathcal{H}&=&B  \psi_{L \downarrow}^\dagger \psi_{R \uparrow} + \Delta \psi_{L \downarrow} \psi_{R \uparrow}+{\rm{H.c.}};
\end{eqnarray}
see Fig. 1. To illustrate the nature of the transition between the gapped phases, we consider
for simplicity the limiting case $\mu=0$ and introduce two Majorana fields $\psi_{L
  \downarrow} (x)=(i \chi_1 (x)+ \chi_2 (x))/\sqrt{2}$, $\psi_{R
  \uparrow}(x)=(\bar{\chi}_1(x) +i \bar{\chi}_2(x))/\sqrt{2}$, in terms of
which the Hamiltonian reads $\mathcal{H}_0 +\delta
\mathcal{H}=\sum_{j=1}^2 \left(\chi_j i\frac{v_F}{2} \partial_x \chi_j
  - \bar{\chi}_j i\frac{v_F}{2} \partial_x \bar{\chi}_j + i m_j \chi_j
  \bar{\chi}_j \right)$, with $m_{1,2}=\Delta \mp B$ (with gauge choice $\Delta>0$). At $|B| =
\Delta$, one of the two Majorana modes becomes massless which is a
signature of a bulk Majorana/Ising quantum phase transition (QPT).~\cite{zuber} Tuning parameters in real space,
so that $m_1(x)$ changes sign, one obtains a Majorana zero mode
localized around $x_0$ with $m_1(x_0)=0$, as can be verified by an
explicit solution of the inhomogeneous quadratic theory. The Majorana edge modes
forming at the interface between distinct phases are topologically
protected and, hence, candidate quantum bits.

Here we will focus on the impact of interactions on the QPT, $B\simeq \Delta$. To see that sufficiently weak
interactions have no qualitative effect, we consider the Majorana
representation of an interaction term, $\mathcal{H}_{{\rm{int}}} \sim
\chi_1 \bar{\chi}_1 \chi_2 \bar{\chi}_2$, and treat the gapped
Majorana component as a c-number $\chi_2 \bar{\chi}_2 \to \langle
\chi_2 \bar{\chi}_2 \rangle$. The ensuing mean field approximation can
be absorbed into a redefined mass parameter $m_1$ of the critical mode,
i.e. a weakly shifted transition point.

\subsection{Generic interactions and mapping to a spin-chain model}
The helical liquid permits two types of time reversal invariant interactions, forward and
umklapp (two-particle) scattering~\cite{Xu,bernevig}
\begin{eqnarray}
\mathcal{H}_{{\rm{fw}}}&=&g_2   \psi_{L \downarrow}^\dagger \psi_{L \downarrow} \psi_{R \uparrow}^\dagger \psi_{R \uparrow}+\frac{g_4}{2}[(\psi_{L \downarrow}^\dagger \psi_{L \downarrow} )^2+(\psi_{R \uparrow}^\dagger \psi_{R \uparrow})^2], \nonumber \\
\mathcal{H}_{{\rm{um}}}&=&g_u  \psi_{L \downarrow}^\dagger \partial_x \psi_{L \downarrow}^\dagger  \psi_{R \uparrow} \partial_x \psi_{R \uparrow}    + {\rm{H.c.}}.
\end{eqnarray}
To make progress with the interacting model, we map the Hamiltonian
$\mathcal{H}=\mathcal{H}_0
+\mathcal{H}_{{\rm{fw}}}+\mathcal{H}_{{\rm{um}}}+\delta \mathcal{H}$
to the XYZ spin
chain model with both non-staggered and staggered magnetic fields, $H_{XYZ} = \sum_{i} {\cal H}_i$ (up to a constant), where
\begin{equation}
\label{xyz}
{\cal H}_i =  \sum_{a=x,y,z}   J_a S^a_i S^a_{i+1} -[\mu+B(-1)^i] S^z_i ,
\end{equation}
and the coupling constants $J_{x,y}\equiv J \pm \Delta >0$, $J = v_F$, and $J_z>0$
are fixed by the condition that after a
Jordan-Wigner re-fermionization~\cite{Giamarchi}
\be
S^z_j =
a^\dagger_j a_j-\frac{1}{2}, ~S^+_j =a_j^\dagger (-1)^j e^{i \pi
  \sum_{l=1}^{j-1} a^\dagger_l a_l}, \nonumber
  \ee  and expansion of the lattice
fermions $a_j$ in terms of the left and right movers, 
\be
a_j \sim e^{i
  \frac{\pi}{2}x} \psi_{R \uparrow}(x)+e^{-i \frac{\pi}{2} x}
\psi_{L\downarrow}(x), \nonumber
\ee 
(with $x=\alpha j$ and lattice constant $\alpha \to 1$), one recovers the starting
Hamiltonian $\mathcal{H}$ with
$J_z=\frac{g_2}{4}=\frac{g_4}{2}=g_u$.
We emphasize that the details introduced by the lattice, such as the above relation between the interaction couplings $g_2$, $g_4$ and $g_u$, are unimportant
for an identification of the morphology of the global phase diagram and the universality of the QPTs;~\cite{foonote2} on the other hand commensurability with the half filled lattice at $\mu=0$ should not be disregarded as a lattice artifact, since it reflects the physical umklapp process which involves the original zero momentum helical modes; see Fig. 1.~\cite{foonoteloss}

\section{Phase diagram}
\label{se:pd} 
\begin{figure}[h] \begin{center} \includegraphics*[width=85mm]{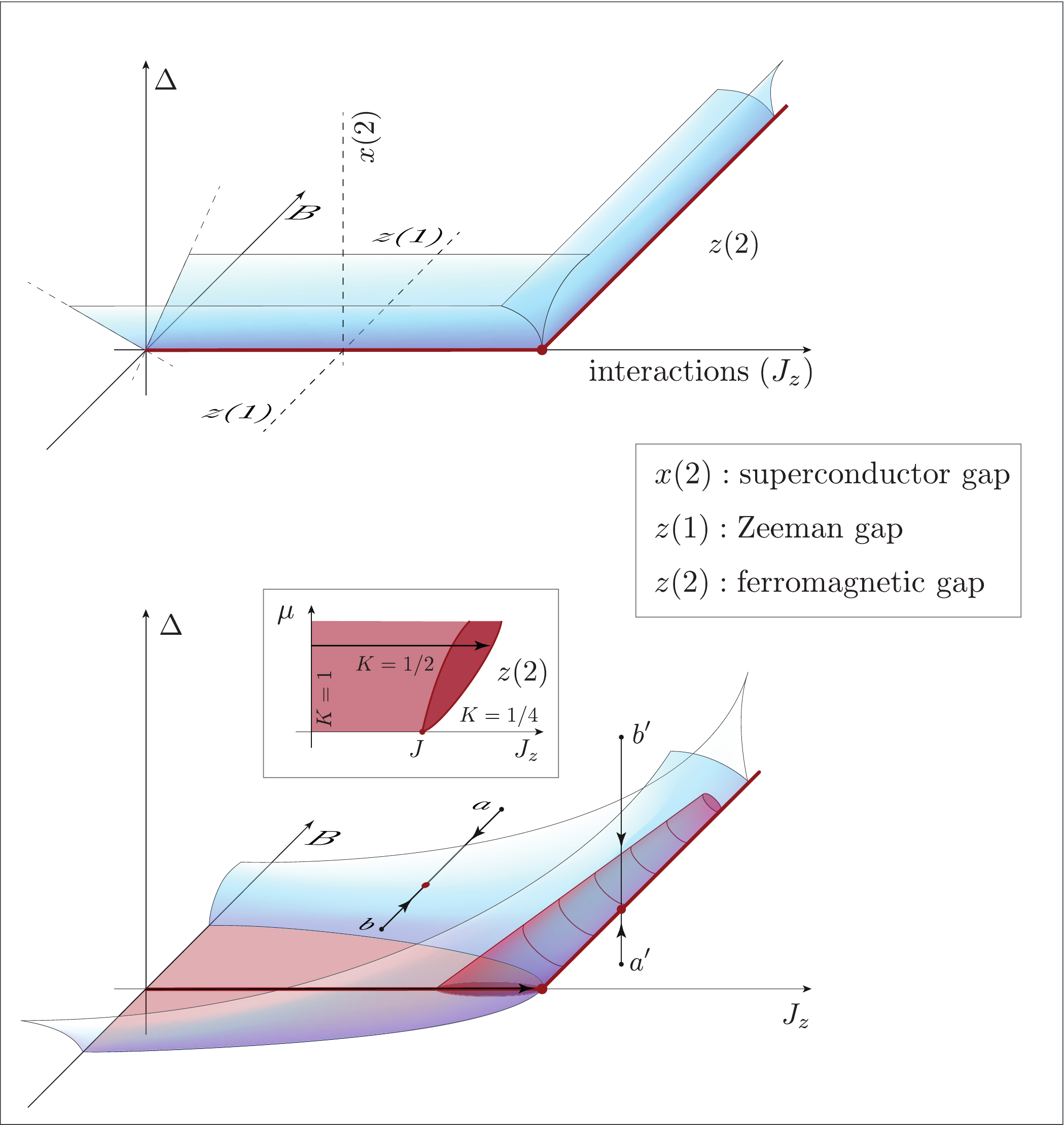}
    \caption{Top: schematic phase diagram of the XYZ model, Eq.~(\ref{xyz}), at
      $\mu =0$. N\'eel ordered phases are denoted by $a(n)$, where
      $a=x,y,z$ denotes the magnetization axis and $n$ the ground
      state degeneracy.  At the thick (red) lines the model reduces to
      a critical XXZ chain described by a LL theory. The blue surfaces
      are Ising transitions described by MF bulk critical modes,
      between the doubly degenerate N\'eel $x$(2) phase and the
      nondegenerate N\'eel $z$(1) phases. The latter are connected with
      the doubly degenerate N\'eel $z$(2) state occurring at $B=0$,
      $J_z>J+|\Delta|$.  Bottom: schematic phase diagram at $\mu \ne 0$.  Inset:
      phase diagram of the XXZ model at $\Delta=B=0$. In the dark shaded region, where $1/2>K>1/4$, the LL phase
      is stable against small $\Delta$.}
\label{fg:1}
\end{center}
\end{figure}

To start our discussion of
topologically distinct phases, we consider the phase diagram spanned
by the parameters, $J_z$, $B$, and $\Delta$, starting with the case
$\mu=0$, cf. Fig.~\ref{fg:1}. In the non-interacting limit, $J_z=0$,
and at the particular parameter configuration $\Delta/J=1$, the
Hamiltonian (\ref{xyz}) is seen to reduce to a transverse field Ising
model (after a $\pi$ rotation of each second spin around $x$),
\begin{equation}
{\cal H}_i \to 2 \Delta S^x_i S^x_{i+1} -B S^z_i .
\end{equation} The latter supports an Ising type transition at $|B|=\Delta$ (blue
diagonal lines in the $J_z=0$ plane of Fig.~\ref{fg:1}): at $|B| < \Delta$ the Ising symmetry $S^x \to - S^x$ is spontaneously broken yielding a doubly degenerate N\'eel-$x(2)$-ordered phase, where $(n)$ denotes a phase with degeneracy $n$. At $|B| > \Delta$ a unique N\'eel-$z(1)$ state is chosen by the direction of the staggered field $B$. The phase transition is long known to have a Majorana
critical mode,~\cite{zuber} as identified above along the $|B|=\Delta$ line. In line with the mean
field argument above, finite $J_z>0$ is irrelevant at this Majorana QPT.

To understand what happens at generic
interaction strength, we first consider the line $\Delta=B=0$
(horizontal red line in Fig.~\ref{fg:1}), where the model reduces to
an XXZ spin chain with spin--$z$ rotation invariance. This model
contains an XXX Heisenberg critical point separating a phase of
gapless planar spin fluctuations at $J_z<J$
[corresponding to a Luttinger liquid (LL) in the fermion
representation] from a doubly degenerate N\'eel-$z(2)$-ordered phase at
$J_z>J$;
expressed in terms of the universal Luttinger interaction parameter,
the transition occurs at $K=1/2$, where $K = 1-\frac{g_2}{2 \pi
  v_F}$ to first order in the  parameters
$g_{2,4,u}$.~\cite{Giamarchi} In fermionic language the order parameter
in the N\'eel-$z(2)$ phase reads $\mathcal{O}=\psi_{L
  \downarrow}^\dagger \psi_{R \uparrow}+ H.c.$, i.e. ferromagnetic
order~\cite{bernevig} (of the \textit{physical} spin) in a direction
parallel to the external field.

Perturbations in both finite $B$ and $\Delta$ around the weak
interaction segment, $1>K>1/2$, are analyzed in the appendix using bosonization and are found to be relevant. Specifically, finite
$\Delta$ perturbes the Hamiltonian by $\delta {\cal H}_i =S^x_i S^x_{i+1} - S^y_i
S^y_{i+1}$, which breaks spin rotation invariance and carries positive
scaling dimension $x_\Delta =2- 1/K$. This operator causes flow towards the doubly
degenerate N\'eel-$x(2)$ order. Finite $B$ couples to the spin chain
through a staggered magnetic field in $z$--direction, leading to an operator with positive scaling dimension $x_B=2-K$, and drives the
system towards the non-degenerate N\'eel-$z(1)$ phase. The discussion
above implies that the line of QPTs $\Delta=B$ at
$J_z=0$ evolves into a `surface' of transition
points (blue surfaces emanating from the segment $J_z/J<1$ with $1/2<K<1$ on the $J_z$ axis in Fig. \ref{fg:1}).
The cusps of these surfaces for $B, \Delta \to 0$,
\bea
\label{cusp}
\Delta_\mathrm{crit} \propto B^{x_\Delta/x_B},
\eea
follow from the difference in scaling
dimensions of the two competing perturbations, $x_B \ge x_\Delta$ for $K \le 1$.

At strong interactions, $K<1/2$, weak $\Delta$ and $B$ become
irrelevant perturbations and cease to affect the N\'eel-$z(2)$ phase.
To understand what happens in this region, it is important to notice
that the Heisenberg point $K=1/2, B=\Delta=0$ is terminal to a second
line of XXZ-models, specified by $J_x=J_z$ or $\Delta=J_z-J\ge0, B=0$
(diagonal red line in the top panel of Fig. \ref{fg:1}). This line
corresponds to gapless planar fluctuations around a conserved
spin-$y$, and its critical properties are equivalent to those of the
horizontal line save for an exchange $z\leftrightarrow y$ of the
invariant axis.~\cite{loss} As follows from a bosonization analysis  of this diagonal line carried out in the appendix, the perturbation $\Delta$  is again relevant with respect to the LL line $\Delta = J_z-J \ge 0$, driving the system either to the N\'eel-$x(2)$
phase for $\Delta > J_z - J$, or to the double
degenerate N\'eel-$z(2)$ ordered phase, for $\Delta < J_z - J$; the
staggered field $B$  being now
perpendicular to the conserved spin-axis is also relevant and
drives the system to the N\'eel-$z(1)$ phase.
We infer that the Majorana transition sheets extrapolate to strong
interactions and merge with this LL as shown in the figure.

\subsection{Phase diagram, ($\mu \ne 0$)}
\label{se:pd1}
At finite $\mu$ the horizontal line $B=\Delta=0$ of gapless
excitations at $J_z<J$ opens to become a surface in the $\Delta=0$
plane. This is easily understood in fermionic language where it means
that for $\mu \ne 0$  a finite magnetic field is needed to create an
excitation gap in the helical liquid as is apparent from the inset of Fig.~1. For larger
interactions the threshold field becomes smaller, which can be seen as
a precursor phenomenon of ferromagnetism. The behavior at the threshold has a universal description in terms of commensurate to incommensurate (C-IC) transitions; see the appendix for more details. At
values of $B$ large enough to sustain a gap, a surface of transition
points $\Delta_\mathrm{crit}=f(B,J_z)$ separates phases of N\'eel $z$
and $x$ order, as qualitatively shown in Fig.~\ref{fg:1} (blueish
surface.)  Importantly, the system supports a second domain of gapless
excitations, shown as a tubular structure in Fig. \ref{fg:1}.

To understand this region, one has to notice that as one sweeps the
interactions in the entire critical region at $B=\Delta=0$ (bolded
line with an end arrow), where Eq. (\ref{xyz}) becomes an exactly solvable XXZ chain in
a longitudinal magnetic field, it is known that $K$ drops from $1$ to $1/4$ at a
C-IC transition (as opposed to $K=1/2$ at the
Kosterlitz-Thouless transition obtained for $\mu=0$, cf.  inset of
Fig. \ref{fg:1}.) This fact entails the
existence of a finite interval of $J_z$ values for which the LL is strongly repulsive $1/2>K>
1/4$ such that $\Delta$ is
irrelevant, $x_\Delta<0$, i.e., there exists a three-dimensional region of parameter
values as indicated in the figure by the red tube for which the system remains gapless, and equivalent to a
Luttinger liquid in the language of fermions. As we argue in the appendix, this critical region is expected to extend to large values of
$\Delta$. Whereas such an interaction induced gapless phase was predicted in the spin polarized quantum wires,~\cite{loss} in the helical liquid considered here upon further increasing interactions the umklapp term becomes relevant at the C-IC transition  at the curve shown as the diagonal red line in the bottom panel in Fig.~\ref{fg:1}, and the gap re-opens. In the appendix we demonstrate that this C-IC transition line connects with the Majorana transition sheets.

\section{Localized states at interfaces}
\label{se:interfaces}
\begin{figure}[h]
  \begin{center} \includegraphics*[width=88mm]{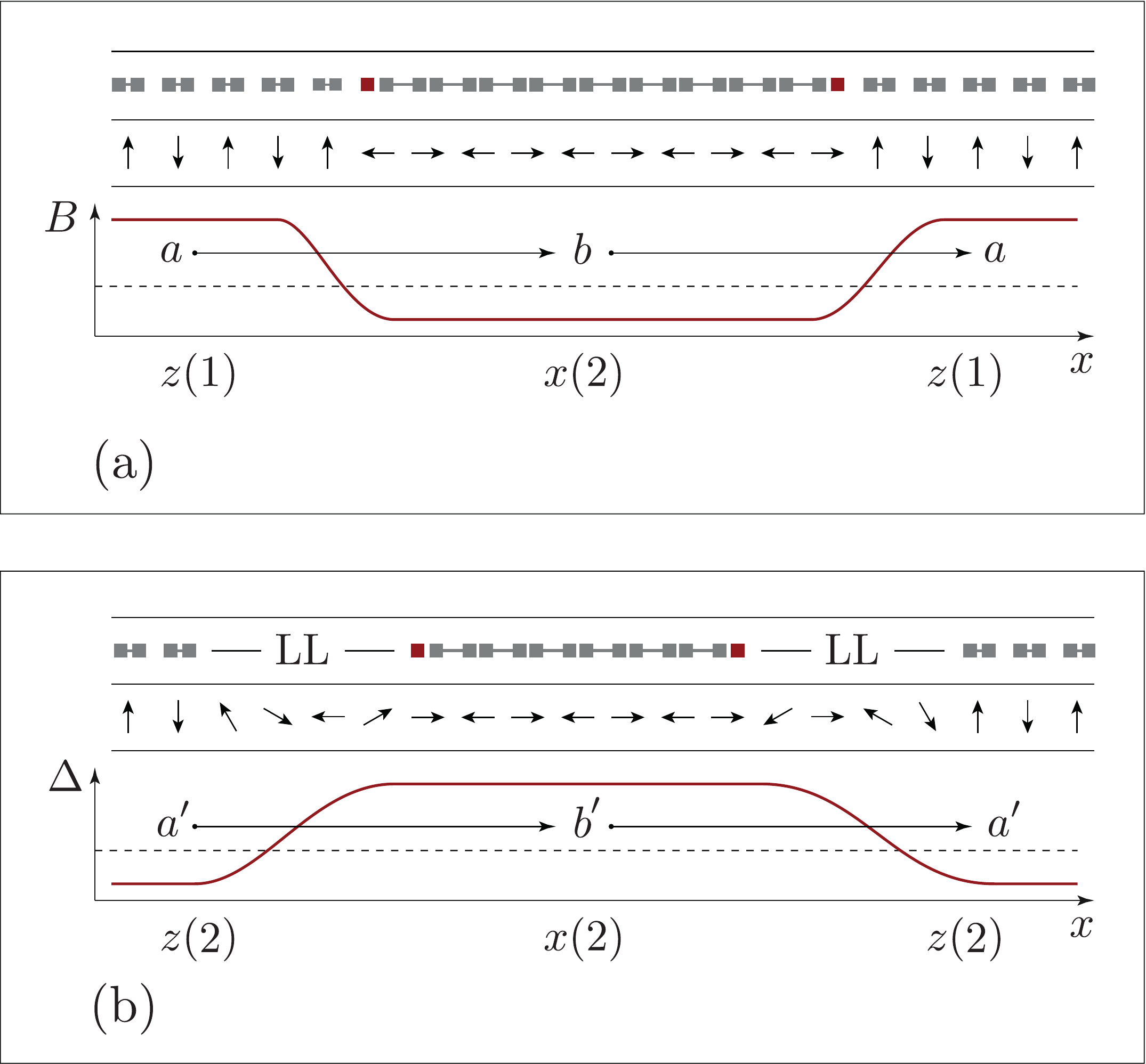} \caption{Spatial
      variation of system parameters at weak (a), or strong (b)
      interactions. Variation profiles correspond to paths $a\to b\to
      a$ or $a'\to b'\to a'$ of Fig.  \ref{fg:1}, respectively. The doubly
      degenerate $x(2)$ center region is
      associated to 2 unpaired MFs (filled red
      squares) located near the interfaces. For strong
      interactions spin
      fluctuating regions form near the interfaces, which in fermion
      language corresponds to  LLs.
     The ensuing low energy modes are coupled to the MFs.\label{fg:1.5}}
\end{center}
\end{figure}

 Having mapped out the phase diagram, we next discuss the implications for
the Majorana edge states. To this end, imagine a space dependent
change of parameters, e.g., along the path $a\to b\to a$ in Fig.
\ref{fg:1}. Within the spin
picture, this leads to a N\'eel $z(1)\to x(2)\to z(1)$ ground state
structure,  cf. Fig. \ref{fg:1.5} upper panel. The central $x(2)$
domain breaking the Ising symmetry $S^x \to - S^x$ is two-fold
degenerate, and at the same time lacks any local order \emph{in fermion language}. This signifies the
presence of unpaired zero energy MFs at the interdaces. This correspondence
can be seen by inspection of the Jordan-Wigner lattice fermion system
corresponding to the center N\'eel $x(2)$ region (sites $1-N$). Introducing a pair of Majorana
fermions for each lattice fermion as $
\gamma_{A,j}=a_j+a_j^\dagger$, and
$\gamma_{B,j}=-i(a_j-a_j^\dagger)$, one finds that the
two terminal  Majorana fermions do not enter the
Hamiltonian.~\cite{kitaev01} The coupling between those MFs,
 \be
 -i \gamma_{A,1} \gamma_{B,N} =(-1)^{N} \sigma^y_1 \left( \prod_{j=2}^{N-1} \sigma^z_j \right) \sigma^y_N,
  \ee
  with Pauli matrices $\sigma_j^a = 2 S_j^a$, is an operator that flips an entire N\'eel $x$ domain, and therefore is suppressed
exponentially in $N$.

More interesting things happen along a sweep of parameters in the strongly interacting system. Consider, for example,
the path $a'\to b'\to a'$ in Fig. \ref{fg:1}, where the proximity
coupling, $\Delta$ is increased in a region of space to pass from a
region with an interaction gap $(a')$ to one with a proximity
gap $(b')$.
This corresponds to an $z(2) \to x(2) \to z(2)$ ground state structure
(Fig. \ref{fg:1.5} bottom) where, however, the $z(2)$ degeneracy of
the outside regions does not imply extra Majorana states
because it corresponds to the degeneracy due to \emph{local} ferromagnetic order in terms of electrons. This shows that the very existence of MFs at interfaces is robust against strong interactions. However, upon moving from $a'$ through the phase boundary towards
$b'$, one crosses the tubular critical region of Fig. \ref{fg:1}. In
spin language, this is a region of gapless planar fluctuations, in
fermion language a LL. Either interpretation shows that
in close \textit{spatial proximity} to the Majorana fermion state a
(Luttinger) liquid of low energy excitations forms. The spatial
extension of this region, $L_\mathrm{LL}$, is the larger the more
shallow the parameter profile of the wire is. While the velocity of excitations vanishes at the C-IC
transition,\cite{Giamarchi} we  nevertheless find by solving an effective Schr\"{o}dinger equation, that
the finite size gap scales as $\sim
v_{\rm eff}/L_\mathrm{LL}$, where $v_{\rm eff}=\mathcal{O}(v_F)$. In comparison, the energy gap towards higher fermionic excitations for weak interactions is much larger and
scales as~\cite{oreg}
%corrected!
$\sim \sqrt{m v_{\rm eff}/L}$  where $m/L$ is the slope of the gap and $L_\mathrm{LL} \sim L$.
Whether or not
fluctuations of the LL get excited primarily depends on the ratio
$T/(v_\mathrm{eff}/L_\mathrm{LL})$ between excitation energies and
temperature, $T$.

\section{Discussion and Summary}
\label{se:discussion}
What is the impact of the LL on the nearby boundary Majorana states?
Unlike non-fermionic low energy quantum fluctuations
(phonons, nuclear spins, etc.), the presence of the LL
may spoil the long term \textit{quantum coherence} of Majorana
fermion states. To
see this, denote the MF operator of a phase crossing
by $\gamma_i$. Non-fermionic excitations do not change the parity of the overall MF number, and in the absence of low energy fermion states, they can couple to MFs only via 
Majorana \textit{bilinears} $i \gamma_i \gamma_j$. In view of the large separation between Majorana sites,
such operators are exponentially suppressed. By contrast, the
coupling to a system of fermionic low energy excitations is mediated
by operators of the form $\gamma_i (c+c^\dagger)$, where $c^\dagger$
is a fermion creation operator. This causes entanglement of
\textit{individual} MFs with a collective quantum
environment, and may compromise long term quantum coherence. It may become an issue at strong interactions in
quantum operations based on the tunnel coupling of endpoint MFs in systems of connected quantum wires.~\cite{alicea} However,  the conservation of the parity of the total number of fermions in the
interface allows to define a protected Majorana subspace,~\cite{akhmerov} which
can be probed by more elaborate interference measurement.~\cite{Hassler}

Summarizing, we have mapped out the phase diagram of interacting
helical liquids subject to both proximity coupling to a superconductor
and a magnetic field. We have reached the
conclusion that even in parameter regions where the interactions are
strong enough to open a gap on their own account, the increase of a
competing proximity gap will generate Majorana excitations. In
general, however, these excitations will suffer decoherence, and this
is due to the fact that in close proximity to the phase boundaries
between interaction and proximity gaps, the system
supports low energy electronic excitations. It will be interesting to
confirm this picture numerically.

\section{Acknowledgements}
We thank A. R. Akhmerov, A. De Martino, V. Gurarie,  F. von Oppen, Y. Oreg, R. G. Pereira and B. Rosenow for important discussions.
This work was
supported by the
A.V. Humboldt Foundation and SFB TR12, FOR 960 and SFB 608 of the DFG.
\appendix

\section{Derivation of the phase diagram from bosonization}

In this appendix we use bosonization to reproduce the phase diagram (PD) in Fig. 2.
The bosonized version of the helical liquid Hamiltonian $\mathcal{H}=\mathcal{H}_0 +\mathcal{H}_{{\rm{fw}}}+\mathcal{H}_{{\rm{um}}}+\delta \mathcal{H}$ given in Eqs.~(1,2,3) reads~\cite{Giamarchi}
\bea
\label{HLLhelical}
\mathcal{H}&=& \frac{v}{2} \left(\frac{1}{K} (\partial_x \phi)^2 +K (\partial_x \theta)^2 \right) -\frac{\mu}{\sqrt{\pi}} \partial_x \phi  \\
&&\!\!\!\! - \frac{g_u}{2 \pi^2 } \cos( \sqrt{16 \pi} \phi)+\frac{B}{\pi } \cos (\sqrt{4 \pi} \phi) -\frac{\Delta}{\pi } \cos ( \sqrt{4 \pi} \theta ), \nonumber
\eea
where $v=v_F+\frac{g_4}{2 \pi} + \mathcal{O}(g_2^2,g_4^2,g_u^2)$ is the renormalized velocity, $K$ is the LL parameter given above to first order in $g_2, g_4, g_u$, the Bose fields satisfy $[\phi(x),\partial_{x'} \theta(x')]=i \delta(x-x')$, and
  the helical electrons are represented by
 $\psi_{R \uparrow(L \downarrow)} \sim \frac{1}{\sqrt{2 \pi }}e^{- i \sqrt{ \pi} (\theta \mp \phi)}$. Eq.~(\ref{HLLhelical}) is also the continuum theory of the XYZ spin chain Eq.~(4), where spins are represented by~\cite{Giamarchi}
\bea
\label{dictionary}
S^z(x)  & \sim &\frac{1}{\sqrt{\pi}} \partial_x \phi(x) + \frac{1}{2 \pi} \cos[\sqrt{4 \pi} \phi(x) - \pi x], \nonumber \\
S^+(x) & \sim &\frac{e^{-i \sqrt{\pi} \theta(x)}}{\sqrt{2 \pi}} \{ (-1)^x+\cos[\sqrt{4 \pi} \phi(x)] \}.
\eea

\subsection{Commensurate case ($\mu=0$)}
 At $B=\Delta=0$ the standard renormalization group  (RG) analysis~\cite{Giamarchi} shows that for infinitesimal $g_u=0^+$ the umklapp term becomes relevant upon increasing interactions beyond a critical value. This occurs at a Kosterlitz-Thouless (KT) transition when $K=1/2$, and  the system goes from a gapless LL at $K>1/2$ to a gapped phase at $K<1/2$ with a pinned value of the bosonic field $\langle \phi \rangle =0$ or $\langle \phi \rangle=\sqrt{\pi}/2$, which classically minimize the $g_u$ term in Eq.~(\ref{HLLhelical}). From Eq.~(\ref{dictionary}), this phase has N\'eel order along $z$,  and corresponds to the N\'eel $z(2)$ phase in the PD at large interaction.

The perturbations $\Delta$ and $B$ around the weak
interaction LL segment, $1>K>1/2$, are both relevant.\cite{Giamarchi} Specifically, infinitesimal
$\Delta$ carries positive
scaling dimension $x_\Delta =2- 1/K$ and leads to pinning of the $\theta$ field at $\langle \theta \rangle=0$ or $\langle \theta \rangle=\sqrt{\pi}$, minimizing the corresponding term in Eq.~(\ref{HLLhelical}). Using Eq.~(\ref{dictionary}), this phase has N\'eel-order along $x$. Analogously, infinitesimal $B>0$  leads to an operator with positive scaling dimension $x_B=2-K$,\cite{Giamarchi} and drives the
system towards a  phase with $\langle \phi \rangle=\sqrt{\pi}/2$ (or $\langle \phi \rangle=0$ for $B<0$), associated with N\'eel-order along $z$. Those two N\'eel $z$ states driven by $B >0$ or $B<0$ are seen to correspond to the two N\'eel $z(2)$ states driven by strong interactions. This implies that the N\'eel $z(1)$ and $z(2)$ phases are connected to each other in the PD without crossing any QPT.
This RG picture valid around the LL segment $1>K>1/2$ (i.e. $0<J<J_z$ in terms of spin chain parameters) implies that the Ising surfaces inferred to emanate from the noninteracting $J_z=0$ plane at $|B| = \Delta$, merge with the LL segment $1>K>1/2$ at $B, \Delta \to 0$.

At strong interactions there is another critical LL line at $\Delta = J_z-J \ge 0$ depicted as the diagonal red line in the top panel in the PD, described by  an XXZ model with $J_x = J_z \ge J_y$ (similarly, there is a third LL line at $\Delta=J-J_z \le 0$ with $J_y=J_z \ge J_x$, not shown). The LL theory is given by $\mathcal{H}_{J_x = J_z \ge J_y} = \frac{v'}{2} \left(\frac{1}{K'} (\partial_x \phi')^2 +K' (\partial_x \theta')^2 \right)$
where the bosonic fields $\phi'$ and $\theta'$ are related to spins by Eq.~(\ref{dictionary}) but with conserved axis being now $y$ rather than $z$, namely
\bea
\label{dictionary1}
S^y  & \sim &\frac{1}{\sqrt{\pi}} \partial_x \phi'+ \frac{1}{2 \pi} \cos(\sqrt{4 \pi} \phi' - \pi x), \nonumber \\
S^z+ i S^x & \sim &\frac{e^{-i \sqrt{\pi} \theta'}}{\sqrt{2 \pi}}[(-1)^x+\cos(\sqrt{4 \pi} \phi')].
\eea
Here, the LL parameter $K'$ and velocity $v'$ are given by~\cite{Giamarchi} $K'=\frac{\pi}{2(\pi-\arccos \frac{J_y}{J_x})}$ and $v'=\frac {J_x \pi}{2} \frac{\sqrt{1-(J_y/J_x)^2}}{\arccos (J_y/J_x)}$.
Different than the $J<J_z$ LL segment, here the staggered field $B$  is perpendicular to the conserved axis, and using Eq.~(\ref{dictionary1}) we find that the most relevant perturbations from this LL line are
\be
\label{LLdiagonal}
\delta H =B \frac{\sqrt{2}}{\sqrt{\pi}} \cos (\sqrt{\pi} \theta' )+\frac{\Delta'}{\pi } \cos ( \sqrt{4 \pi} \theta' ),
\ee
where $\Delta'=\Delta - J_z+J$. The second term, as the term $\propto \Delta$ in Eq.~(\ref{HLLhelical}), is the leading U(1) breaking perturbation  caused by finite $\Delta'$ which renders $J_x \ne J_z$. The scaling dimensions of the two operators in Eq.~(\ref{LLdiagonal}) are $x_B' =2- 1/(4K')$ and $x_\Delta' = 2-1/K'$, respectively.\cite{Giamarchi}
Both operators in Eq.~(\ref{LLdiagonal}) are relevant long the $\Delta = J_z-J \ge 0$ line, tending to localize the $\theta'$ field. They may or may not compete with each other depending on their relative sign. Positive $\Delta'$ locks $\theta'$ to $\langle \theta' \rangle =\frac{ \sqrt{\pi} }{2}$ or $\langle \theta' \rangle =3\frac{ \sqrt{\pi} }{2}$, corresponding to the N\'eel $x(2)$ phase, whereas negative $\Delta'$ locks $\theta'$ to $\langle \theta' \rangle  = 0$ or $\langle \theta' \rangle  = \sqrt{\pi}$, corresponding to the N\'eel $z(2)$ phase. Upon increasing $B$ at fixed $\Delta'<0$ [starting from the $z(2)$ phase below the diagonal red line PD] there is no transition since the minimas of both cosines in Eq.~(\ref{LLdiagonal}) overlap. On the other hand, starting at fixed $\Delta'>0$ [at the $x(2)$ phase above the diagonal red line in the PD] there is a transition upon increasing $B$ to either $\langle \theta' \rangle =0$ for $B<0$ or $\langle \theta' \rangle =\sqrt{\pi}$ for $B>0$, corresponding to the N\'eel $z(1)$ states. Such transition has Ising character.~\cite{schultz,Delfino} This RG picture implies that the Ising surfaces discussed above should eventually merge with the LL line  $\Delta = J_z-J>0$. Close to this line the competing operators with scaling dimensions $x_B' >x_\Delta'$ lead to the cusp singularity $\Delta'_{crit} \propto B^{x_\Delta' / x_B'}$ which connects to Eq.~(\ref{cusp}) at the Heisenberg point $J_x=J_y=J_z$.

\subsection{Incommensurate case ($\mu \ne 0$)} We initially consider the $\Delta=0$ plane, and fix $B$ and $J_z$. Large enough $\mu$, acting as a uniform magnetic field  on the spin chain, will lead to a finite magnetization, $ \langle S^z \rangle \ne 0$. From Eq.~(\ref{dictionary}), this means that the phase $\phi$ will acquire a linear term $\phi(x) =\sqrt{\pi} \langle S^z \rangle x + ...$, where $...$ denote fluctuations, therefore the $B$ and $g_u$ terms in Eq.~(\ref{HLLhelical}) become oscillating in space and can not open a gap. Thus the system is in an incommensurate LL phase in which $ \langle S^z \rangle $ changes continuously as function of $\mu$. Upon decreasing $\mu$ the magnetization decreases till $\langle S^z \rangle \to 0$, at which point the $B$ and $g_u$ terms no longer oscillate, and the system enters into either one of the commensurate $z(1)$ or $z(2)$ phase for either $B\ne0$, or for $B=0$ and sufficiently large $J_z$, respectively. This follows from the stability of the gapped phases in the $\mu=0$ PD against finite $\mu$. Hence there is  a  commensurate to incommensurate (C-IC) transition between the $z(1)$ or $z(2)$ phases and the  LL gapless state as function of $\mu$. Alternatively, at fixed $\mu$ this transition occurs as function of $B$ or $J_z$ as shown in the PD. The value of the LL parameter at a C-IC transition is renormalized to the universal value $K=1/n^2$,\cite{Giamarchi} where $n$ is the degeneracy of the commensurate phase. Hence, we infer the existence a critical LL surface in the $\Delta=0$ plane in the PD at $\mu \ne 0$, with LL parameter  $K=1$ at its edges, except at the $B=0$ tip, where $K=1/4$.

\begin{figure}[h]
  \begin{center} \includegraphics*[width=40mm]{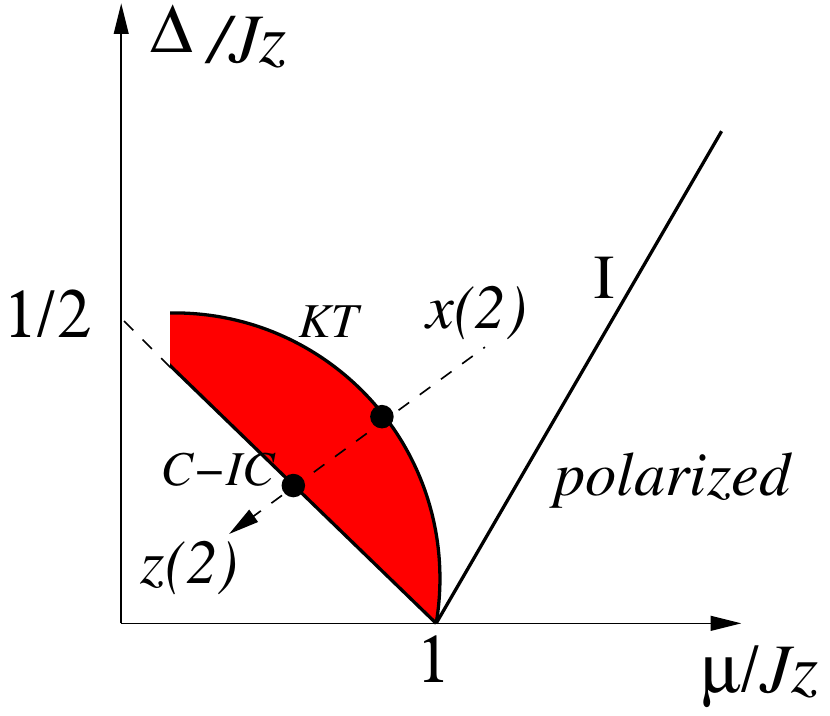} \caption{Schematic phase diagram of the ANNNI model Eq.~(\ref{ANNNI}). Sarting at the N\'eel-x(2) phase, upon increasing the interaction $J_z$ at fixed $\Delta /\mu$ (trajectory along the dashed line) one first crosses a KT  transition into a gapless LL phase, and then one crosses a C-IC transition into the N\'eel-z(2) phase. This trajectory corresponds to crossing the red tube in Fig. 2 at $\Delta=J$, $B=0$, $\mu \ne 0$, upon increasing $J_z$. The ANNNI model also has an Ising (I) transiton between the N\'eel-x(2) phase and a polarized phase.  \label{fg:annni}}
\end{center}
\end{figure}

Using the field theory Eq.~(\ref{HLLhelical}) describing the critical surface at the $\Delta=0$ plane, we may analyze the effect of an infinitesimal $\Delta$. Since $x_\Delta = 2-1/K$, this perturbation is relevant for $K>1/2$, and in this case the $\theta$ field is pinned and the system enters the N\'eel-$x(2)$ phase for infinitesimal $\Delta$. However $\Delta$ is irrelevant for $K<1/2$. Since $K$ varies continuously in the LL surface at $\Delta=0$ except at the the $B=0$ tip with $K=1/4$, we conclude that there must exist  a  finite critical subarea with $K<1/2$, as shown schematically by the red dark area in the $\Delta=0$ plane in the bottom panel of the PD. This critical subarea is stable against $\Delta$ and hence extends into a three dimensional critical region above the $\Delta=0$ plane up to some finite value of $\Delta$, as shown schematically by the tube in the PD; as follows from the universal values of the LL parameter at C-IC transitions quoted above, we conclude that $K=1/2$ at the 2D boundary of this tube with the N\'eel $x(2)$ phase, and $K=1/4$ at its C-IC transition with the  N\'eel $z(2)$ phase. At $B=0$, the C-IC transition is the only region in the PD which is sensitive to small $B$. Hence, the Majorana transition surfaces must merge with this C-IC transition line.

It is natural to expect that this 3D manifold extends to large
values of $\Delta$ along the critical line $\Delta=J_z-J$ (for $\mu \to 0$ the critical manifold will shrink to this critical line). As an evidence for this expectation, we observe that for $B=0$ and $\Delta=J$  Eq.~4 reduces to the extensively studied~\cite{review} ANNNI (Axial Next Nearest Neighbor Ising) model,
\be
\label{ANNNI}
{\cal H}_i \to 2 \Delta S^x_i S^x_{i+1}+J_z  S^z_i S^z_{i+1} -\mu S_i^z,
\ee
 whose phase diagram is shown schematically  in Fig.~\ref{fg:annni}. The so-called ``floating phase''~\cite{review} of this model (red shaded area), can be identified with the LL phase inside the critical tube in Fig. 2 at finite $\mu$. Note, however, that whereas there exists an analytic proof that this LL phase emerges out of the $\Delta=0$, $\mu= J_z$ multicritical point in Fig.~\ref{fg:annni},~\cite{Villain,Rujan} there is still no full evidence that this LL phase extends to the $x(2)$-$z$(2) transition point at $\mu=0$.~\cite{Allen,feo}


\begin{thebibliography}{99}
\bibitem{foonote2} In quantum wires $g_u$ is typically smaller than $g_{2,4}$ as it is generated by a combination of interaction and spin-orbit coupling.
\end{thebibliography}
\end{document}